\documentstyle[aps,prl,floats,graphicx,epsfig]{revtex}
\begin{document}
\draft
\twocolumn[\hsize\textwidth\columnwidth\hsize\csname @twocolumnfalse\endcsname

\title{Vortex Mass in BCS systems: Kopnin and Baym-Chandler
contributions. }
\author{  G.E. Volovik}

\address{Low Temperature Laboratory, Helsinki University of
Technology, P.O. Box 2200, FIN-02015 HUT, Finland, \\ and \\ Landau
Instute for  Theoretical
Physics,
117334
Moscow, Russia}

\date{\today}
\maketitle

\begin{abstract}

The Kopnin mass and the Baym-Chandler mass of the vortex have the
same origin. Both represent the mass of the normal component
trapped by the vortex. The Kopnin mass of the vortex is formed by
quasiparticles localized in the vicinity of the vortex. In the
superclean limit it is calculated as linear response, exactly in the
same way as the density of the normal component is calculated in
homogeneous superfluid. The  Baym-Chandler mass is the hydrodynamical
(associated) mass trapped by vortex. It is analogous to the normal
component formed by inhomogeneities, such as pores and impurities.
Both contributions are calculated for the generic model of the
continuous vortex core.

\end{abstract}
\

\

]
\narrowtext

It is well known that in the BCS superfluids and superconductors
the most important contribution to the vortex mass originates from
the vortex core. The core mass in these systems is proportional
to the area of the core $A\sim \xi^2$, where  $\xi$ is
the coherence length  (see \cite{Kopnin1978}
for the vortex mass  in superconductors and \cite{KopninSalomaa} for
the vortex mass in superfluid $^3$He-B). This core mass is essentially larger
than the logarithmically divergent contribution, which comes from the
compressibility. Inspite of the logarithmic divergence, the latter
contains the speed of sound in the denominator and thus is smaller
by factor
$(a/\xi)^2 \ll 1$,  where $a$ is the interatomic distance. The
compressibility mass of the vortex dominates in Bose superfluids,
where the core size is small, $\xi
\sim a$.

According to Kopnin theory the core mass comes from the fermions
trapped in the vortex core
\cite{Kopnin1978,KopninSalomaa,vanOtterlo1995,Volovik1996VortexMass}.
Recently the problem of another vortex mass of the hydrodynamics
origin was raised in Ref.\cite{Sonin}. It is the so-called backflow
mass discussed by Baym and Chandler \cite{BaymChandler}, which also
can be proportional to the core area.  Here we compare these two
contributions in the superclean regime and at low $T\ll T_c$ using
the model of the continuous core.

The continuous-core vortex is one of the best models which
helps to solve many problems in the vortex core physics. Instead of
consideration of the singular core, one can smooth the
$1/r$-singularity of the superfliuid velocity  by introducing the
point gap nodes in the core region. As a result the
superfluid/superconducting state in the vortex core of any system
acquires the properties of the A-phase of superfluid $^3$He
with its continuous vorticity and point gap nodes
\cite{VolovikMineev1982,SalomaaVolovik}.
Using the continuous-core model one can show for example that the
Kopnin (spectral flow) force comes directly from
the Adler-Bell-Jackiw chiral anomaly equation \cite{Bevan} and this
shows the real origin of this anomalous force. In this model one
can easily separate different contributions to the vortex mass.
Actually this is not only the model: The spontaneous smoothing of the
velocity singularity occurs in the core of both types of vortices
observed in $^3$He-B
\cite{SalomaaVolovik}; in heavy fermionic and high-$T_c$
superconductors such smoothing can occur due to admixture of
different pairing states in the vortex core.

It appears that both the Kopnin mass and the Baym-Chandler mass
are related to the normal component. In general the normal component
of the superfluid liquid comes from two sources: (i) the
local contribution, which comes from the system of quasiparticles,
and (ii) the associated mass related to the backflow.  The latter is
dominating in porous materials, where some part of superfluid
component  is hydrodynamically trapped by the pores and thus is
removed from the overall superfluid motion. The normal component,
which gives rise to the vortex mass contains precisely the same
two contributions. (i) The local contribution  comes from the
quasiparticles localized in the vortex core and thus moving with the
core. This is the origin of the Kopnin mass according to
Ref.\cite{Volovik1996VortexMass}. (ii) The associated mass
contribution arises because the profile of the local density of the
normal component in the vicinity of the vortex core  disturbs the
superflow around the vortex, when the vortex moves. This creates the
backflow and thus some part of the superfluid component is trapped
by the moving vortex, resulting in the Baym-Chandler mass of the
vortex.

\begin{figure}[!!!t]
\begin{center}
\leavevmode
\epsfig{file=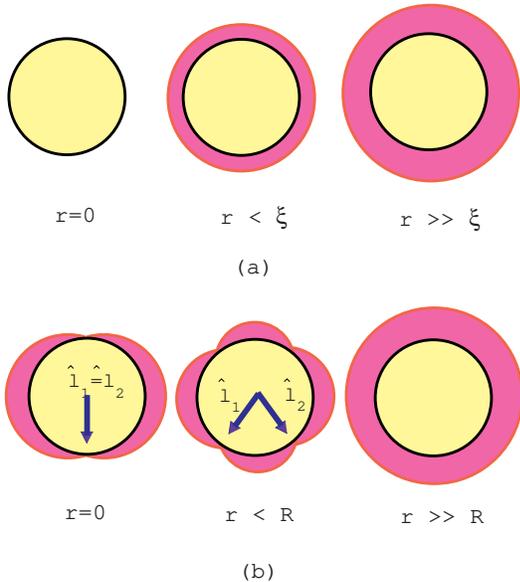,width=0.8\linewidth}
\caption[Core]
    {Singular vortex vs soft-core vortex. (a) In the singular
vortex the gap continuously decreases and becomes zero on the
vortex axis (at
$r=0$) (b) For some vortices it is energetically favourable to
escape the nullification of the order parameter at $r=0$. Instead, within the
smooth core, $r<R$, the point gap nodes appear in the spectrum of fermions
\cite{VolovikMineev1982}. The unit vectors $\hat{\bf l}_1$ and
$\hat{\bf l}_2$ show the directions to the nodes at different
$r$. Close to the gap nodes the spectrum of fermions is similar to
that in
$^3$He-A. }
\label{Core}
\end{center}
\end{figure}

Let us consider this on the example of the simplest continuous-core
vortex (Fig.\ref{Core}). It has the following distribution of the unit vector
$\hat {\bf l}({\bf r})$ which shows the direction of the pont gap nodes in
the smooth core
\begin{equation}
\hat {\bf l}({\bf r})=\hat {\bf z}\cos\eta(r) + \hat {\bf
r}\sin\eta(r) ~~,
\label{l}
\end{equation}
where $z,r,\phi$ are cylindrical coordinates. For superfluid $^3$He-A
the $\hat {\bf l}$-vector in the smooth core changes from $\hat {\bf
l}(0)=-\hat {\bf z}$ to $\hat {\bf l}(\infty)=\hat {\bf z}$, which
represents the doubly quantized continuous vortex. For the
smoothed singly quantized vortices in $^3$He-B and superconductors
one has two $\hat {\bf
l}$-vectors with  $\hat
{\bf l}_1(0)=\hat
{\bf l}_2(0)=-\hat {\bf z}$ and $\hat {\bf l}_1(\infty)=-\hat
{\bf l}_2(\infty)=\hat {\bf r}$ \cite{SalomaaVolovik}. The region
of radius $R$, where the texture of $\hat{\bf l}$-vectors is
concentrated, is called the smooth (or soft) core of the vortex.

{\it Kopnin mass}. Let us remind the phenomenological derivation of
the Kopnin mass of the vortex at low $T$ and in the
superclean regime \cite{Volovik1996VortexMass}. If the
vortex moves with  velocity
${\bf v}_L$ with respect to the superfluid component, the fermionic
energy spectrum in the vortex frame is Doppler shifted:
$E=E_0(\nu)- {\bf k}\cdot  {\bf v}_L$, where $\nu$ stands for the
fermionic degrees of freedom in the stationary vortex.  The summation
over  fermionic degrees of freedom  leads to
the extra linear momentum of the vortex
$\propto{\bf v}_L$:
\begin{equation}
{\bf P}=\sum_\nu {\bf k}\theta (-E)=\sum_\nu {\bf k} ({\bf k}\cdot
{\bf v}_L) \delta (E_0) =M_{\rm Kopnin}{\bf v}_L ~~,
\label{DOS}
\end{equation}
Note that this vortex mass is determined in essentially the same
way as the normal component density in the bulk system. The Kopnin
vortex mass is nonzero if the density of fermionic states is finite
in the vortex core. The density of states (DOS) is determined by the
interlevel spacing
$\omega_0$ in the core:
$N(0)\propto 1/\omega_0$, which gives for the Kopnin vortex mass an
estimation:
$M_{\rm Kopnin}
\sim k_F^3/\omega_0$ (The exact expession is $M_{\rm Kopnin}
=  \int_{-k_F}^{k_F} (dk_z/
4\pi)(k_\perp^2  /
\omega_0(k_z))  $). For the soft-core vortex the interlevel spacing
is $\omega_0\sim
\hbar^2/(m\xi R)$ which gives the Kopnin mass $M_{\rm Kopnin}
\sim\rho \xi R$,\cite{Kopnin1995} where $\rho$ is the mass density
of the liquid.

For our purposes it is instructive to consider the normal component
associated with the vortex as the local quantity, determined at
each point in the vortex core. Such consideration is valid for the
smooth core with the radius $R\gg \xi$, where the local classical
description of the fermionic spectrum can be applied. The main
contribution comes from the point gap nodes, where the classical
spectrum has the form $E_0=\sqrt{v_F^2(k-k_F)^2 +
\Delta_0^2(\hat{\bf k}\times \hat {\bf l})^2}$ and  $\Delta_0$ is
the gap amplitude. In the presence of the gradient of $\hat {\bf
l}$-field, which acts on the quasipaerticles as an effective
magnetic field, this  gapless spectrum leads to the nonzero local DOS
and finally to the following local density of the normal component at
$T=0$ (see Eq.(5.24) in the review \cite{Volovik1990}):
\begin{equation}
(\rho_n)_{ij}({\bf r})=\rho_n \hat  l_i\hat
l_j~,~\rho_n={k_F^4 \over 2\pi^2 \Delta_0}
\vert (\hat {\bf l}\cdot \vec\nabla)\hat {\bf l}\vert ={k_F^4
\sin\eta\over 2\pi^2 \Delta_0}
|\partial_r \eta|.
\label{NormalComponent}
\end{equation}

The integral of this normal density tensor over the cross section of
the soft core gives the same Kopnin mass of the vortex but in
the local density representation \cite{Volovik1996VortexMass}
\begin{equation}
M_{\rm Kopnin}= {1\over 2} \int d^2r \rho_n (r) \sin^2\eta(r) \sim
 \rho \xi R ~.
\label{KopninMass}
\end{equation}
He we used that $v_F/\Delta_0 \sim \xi$. Note that area law for the
vortex mass is valid only for vortices with
$R\sim \xi$, but in general one has the linear law: $M_{\rm Kopnin}
\sim \rho \xi R$. \cite{Kopnin1995}

{\it The associated (or induced) mass} appears when, say, an
external body moves in the superfluid. This mass depends on the
geometry of the body. For the moving cylinder of radius $R$ it is
the mass of the liquid displaced by the cylinder,
\begin{equation}
M_{\rm associated}=\pi R^2
\rho ~~,
\label{AssociatedMassCylinder}
\end{equation}
which is to be added to the actual mass of the cylinder to obtain the
total inertial mass of the body. In superfluids this part
of superfluid component moves with external body and thus can be
associated with the normal component.  The similar mass
is responsible for the normal component in porous materials and in
aerogel, where some part of superfluid  is hydrodynamically trapped by
the pores. It is removed from the overall superfluid motion and
thus becomes the part of the normal component.

In the case when  the vortex is trapped
by the wire of radius $R\gg \xi$,  such that the vortex core is
represented by the wire, the Eq.(\ref{AssociatedMassCylinder}) gives
the vortex mass due to the backflow around the moving core.  This is
the simplest realization of the   backflow mass of the vortex
discussed by Baym and Chandler
\cite{BaymChandler}.  For such vortex with the wire-core the
Baym-Chandler mass is the dominating mass of the vortex. The
Kopnin mass which can result from the normal excitations trapped
near the surface of the wire is essentially less.

Let us now consider the Baym-Chandler mass for the free vortex at
$T=0$ using again the continuous-core model. In the wire-core vortex
this mass arises due to the backflow caused by the inhomogeneity of
$\rho_s$: $\rho_s(r>R)=\rho$ and $\rho_s(r<R)=0$. Similar but less
severe inhomogeneity of $\rho_s=\rho -\rho_n$ occurs in the
continuous-core vortex due to the nonzero local normal density in
Eq.(\ref{NormalComponent}). Due to the profile of the
local superfluid density the external flow  is disturbed
near the core according to continuity equation
\begin{equation}
{\bf \nabla}\cdot (\rho_s {\bf v}_s)=0 ~~.
\label{ContinuityEquation1}
\end{equation}
If the smooth core is large, $R\gg \xi$, the deviation  of the
superfluid component in the smooth core from its asymptotic value
outside the core is small:
$\delta \rho_s=\rho -\rho_s \sim (\xi/R) \rho \ll \rho$ and can be
considered as perturbation. Thus if the asymptotic value of the
velocity  of the superfluid component with respect of the core is
${\bf v}_{s0}=-{\bf v}_L$, the disturbance $\delta {\bf
v}_{s}={\bf \nabla}\Phi$ of the superflow in the smooth core is
given by:
\begin{equation}
\rho {\bf \nabla}^2 \Phi=  v_{s0}^i\nabla^j (\rho_n)_{ij} ~~.
\label{ContinuityEquation2}
\end{equation}
The kinetic  energy of the backflow gives the  Baym-Chandler mass
of the vortex
\begin{equation}
M_{\rm BC}={\rho \over  v_{s0}^2} \int d^2r \left({\bf \nabla}  \Phi
\right)^2 ~~,
\label{BCmass1}
\end{equation}
In the simple approximation, when the normal
component in Eq.(\ref{NormalComponent}) is considered as
isotropic,   one obtains
\begin{equation}
M_{\rm BC}={1\over 2 \rho} \int d^2r \rho_n^2(r) \sim \rho \xi^2 ~~.
\label{BCmass2}
\end{equation}

The Baym-Chandler mass
does not depend on the core radius $R$, since the
large area $R^2$ of integration in Eq.(\ref{BCmass2}) is
compensated by small value of the normal component in the rarified
core, $\rho_n \sim \rho (\xi/R)$. That is why  if the
smooth core is large, $R\gg \xi$,    this mass is   parametrically
smaller than the Kopnin mass in Eq.(\ref{KopninMass}).

In conclusion, both contributions to the mass of the vortex result
from the mass of the normal component trapped by the vortex.  The
difference between Kopnin mass and Baym-Chandler backflow mass  is
only in the origin of the normal component trapped by the vortex.
The relative importance of two masses depends on the vortex core
structure: (1) For the free continuous vortex with the large core
size $R\gg \xi$, the Kopnin mass dominates:
$M_{\rm Kopnin} \sim \rho R\xi \gg  M_{\rm BC} \sim \rho \xi^2$.
(2) For the vortex trapped by the wire of radius $R\gg \xi$,
the Baym-Chandler mass is proportional to the core area, $M_{\rm BC}
\sim \rho R^2$, and is parametrically larger than the Kopnin mass.
(3) For the free vortex core with the core radius $R\sim \xi$ the
situation is not clear since the continuous core approximation does
not work any more. But extrapolation of the result in
Eq.(\ref{BCmass2}) to
$R\sim \xi$ suggests that the Baym-Chandler mass   can be
comparable with Kopnin mass.

I thank G. Blatter and N.B. Kopnin  for numerous discussions.
This work was supported by the Russian Foundation for Fundamental
Research grant No. 96-02-16072, by the RAS program ``Statistical
Physics'' and by the Intas grant 96-0610.


\begin{references}


\bibitem{Kopnin1978} N.B. Kopnin,  Pis'ma
ZhETF, {\bf 27}, 417 (1978); \lbrack   JETP Lett., {\bf 27}, 390
(1978) \rbrack.

\bibitem{KopninSalomaa}  N.B. Kopnin and M.M. Salomaa, Phys. Rev.
{\bf  B~44}, 9667  (1991).

\bibitem{vanOtterlo1995} A. van Otterlo, M.V. Feigel'man, V.B.
Geshkenbein, and G. Blatter, Phys. Rev. Lett. {\bf 75}, 3736 (1995).

\bibitem{Volovik1996VortexMass} G.E. Volovik,     Pis'ma ZhETF {\bf
65}, 201  (1997)   \lbrack  JETP Lett. {\bf 65 },  217
(1997)\rbrack .

\bibitem{Sonin} E.B. Sonin,  V.H. Geshkenbein, A. van Otterlo
and G. Blatter, Phys. Rev.      {\bf B~57}, 575 (1998).

\bibitem{BaymChandler}  G. Baym and E. Chandler, J. Low Temp.
Phys.   {\bf 50}, 57  (1983).

\bibitem{VolovikMineev1982} G.E. Volovik and V.P. Mineev,
ZhETF {\bf 83}, 1025  (1982) \lbrack JETP {\bf 56},  579
(1982)\rbrack.

\bibitem{SalomaaVolovik}  M.M. Salomaa and  G.E. Volovik,
Rev. Mod. Phys. {\bf 59}, 533  (1987)

\bibitem{Bevan} T.D.C. Bevan, A.J. Manninen, J.B. Cook, J.R.
Hook, H.E. Hall, T. Vachaspati and G.E. Volovik,    Nature,  {\bf
386},  689   (1997).

\bibitem{Kopnin1995} N.B. Kopnin, Physica {\bf B~ 210}, 267 (1995).

\bibitem{Volovik1990} G.E. Volovik, in: {\bf Helium Three}, eds.
W.P.Halperin, L.P.Pitaevskii, Elsevier Science Publishers B.V., p.
27,  1990.


\end{references}
\end{document}